\documentclass[runningheads]{llncs}
\usepackage[misc]{ifsym}
\usepackage{bbding}
\usepackage{amsmath}
\usepackage{graphicx}
\usepackage{multirow}
\usepackage{array}
\usepackage[colorlinks,linkcolor=blue,anchorcolor=blue,citecolor=blue]{hyperref}

\usepackage{float}
\usepackage{multicol}
\begin{document}
\title{CS$^2$: A Controllable and Simultaneous Synthesizer of Images and Annotations with Minimal Human Intervention}
\titlerunning{CS$^2$: A Controllable and Simultaneous Synthesizer}
%
%
\author{Xiaodan Xing\inst{1} \and
Jiahao Huang\inst{1, 5}\and
Yang Nan\inst{1}\and
Yinzhe Wu\inst{1, 2}\and
Chenjia Wang\inst{3}\and
Zhifan Gao\inst{4}\and
Simon Walsh \inst{1,*}  \and
Guang Yang\inst{1, 5,*} \Envelope}

\authorrunning{Xing et al.}

\institute{National Heart and Lung Institute, Imperial College London, London, UK \\
\email{g.yang@imperial.ac.uk}
\and
Department of Biomedical and Engineering, Imperial College London, London, UK
\and
Edinburgh Centre for Robotics, Heriot-Watt University, Edinburgh, UK \and
School of Biomedical Engineering, Sun Yat-sen University, Guangdong, China\and
Cardiovascular Research Centre, Royal Brompton Hospital, London, UK\\
\email{*Co-last senior authors}
\vspace{-1em}}
\maketitle              
\begin{abstract}
The destitution of image data and corresponding expert annotations limit the training capacities of AI diagnostic models and potentially inhibit their performance. To address such a problem of data and label scarcity, generative models have been developed to augment the training datasets. Previously proposed generative models usually require manually adjusted annotations (e.g., segmentation masks) or need pre-labeling. However, studies have found that these pre-labeling based methods can induce hallucinating artifacts, which might mislead the downstream clinical tasks, while manual adjustment could be onerous and subjective. To avoid manual adjustment and pre-labeling, we propose a novel controllable and simultaneous synthesizer (dubbed CS$^2$) in this study to generate both realistic images and corresponding annotations at the same time. Our CS$^2$ model is trained and validated using high resolution CT (HRCT) data collected from COVID-19 patients to realize an efficient infections segmentation with minimal human intervention. Our contributions include 1) a conditional image synthesis network that receives both style information from reference CT images and structural information from unsupervised segmentation masks, and 2) a corresponding segmentation mask synthesis network to automatically segment these synthesized images simultaneously. Our experimental studies on HRCT scans collected from COVID-19 patients demonstrate that our CS$^2$ model can lead to realistic synthesized datasets and promising segmentation results of COVID infections compared to the state-of-the-art nnUNet trained and fine-tuned in a fully supervised manner. 
\end{abstract}
\section{Introduction}
\par{Medical images with ample annotations are difficult to obtain. This paucity of medical images and labels compromises the performance of AI-based computer-aided diagnosis, including medical image segmentation and classification. Medical image synthesis \cite{burgos2020simulation,dalmaz2021resvit} provides an effective and practical strategy for data augmentation that can mitigate the problem of data and labeling scarcity. Generative Adversarial Networks (GAN) \cite{goodfellow2014generative} is a potent paradigm for realistic medical image synthesis, which can be potentially used for effective data augmentation. In medical image synthesis, three types of GANs are widely used:}
\par
\begin{figure}
\includegraphics[width=\textwidth]{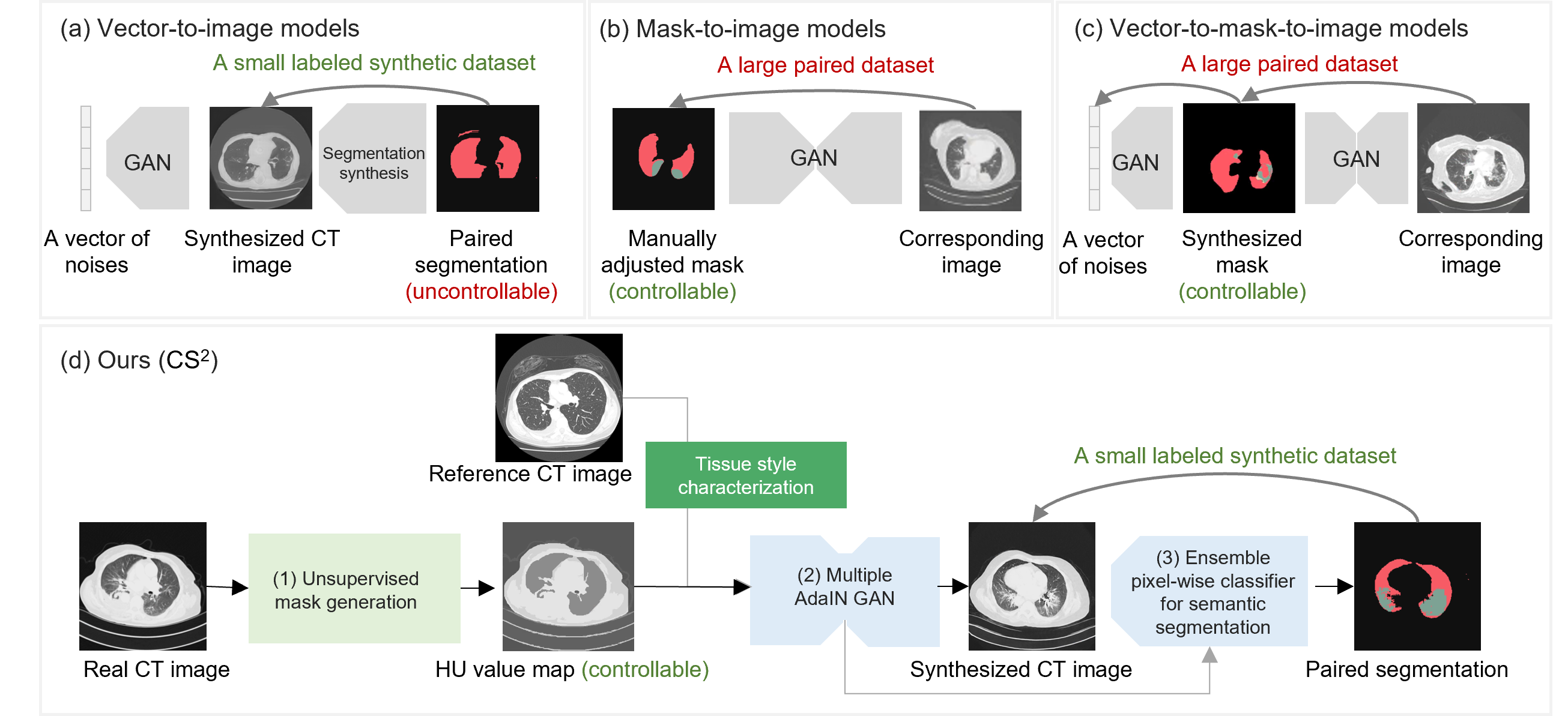}\vspace{-1em}
\caption {The rationale of our developed CS$^2$ compared to previously proposed data augmentation methods, including vector-to-image (a), mask-to-image (b), and vector-to-mask-to-image (c). Our CS$^2$ model can produce a large labeled synthetic dataset controllably without using large-scale pre-labeling or human-labeled datasets.\vspace{-1em}} \label{fig:1}
\end{figure}

{The first type of model is called vector-to-image (V2I) GANs, which synthesize augmented images from a vector of noises (Fig. \ref{fig:1} (a)). A recent study \cite{zhang2021datasetgan} used a V2I styleGAN model to produce annotated augmented images from a vector of noises without a pre-labeled dataset. By labeling a handful of those synthesized images and re-inferencing their styleGAN, augmented images with annotations could be obtained. However, the generation of anatomy and the location of the lesions could not be controlled in these V2I models, and the explainability of these generative models was limited, which might sabotage following clinical interpretation and decision making.}
\par{The second type is called mask-to-image (M2I) GANs (Fig. \ref{fig:1} (b)). The annotations of these algorithms could be manually adjusted, i.e., combining existing segmentation masks \cite{shin2018medical,sun2020mm}, or transferred from a labeled dataset in another domain \cite{huo2018adversarial}. Unfortunately, fully paired mask-to-image synthesis reduces the variance of synthetic data. Moreover, these models require large human input to feed them with manual pre-labeling, conflicting with our primary aim to alleviate manual labeling workload. Besides, cross-modality models require the labeled dataset in another domain, which might not always be available due to exorbitant costs of additional scanning and patients’ physiological limitations.}  
\par{The third type is vector-to-mask-to-image (V2M2I) GANs. To obtain the segmentation mask, a particular synthesis network in addition to the networks in M2I was designed and trained \cite{bailo2019red,pandey2020image}. However, the added segmentation mask synthesis network demanded a large number of labeled samples to make it well trained, which could be up to hundreds for example in \cite{bailo2019red} thus labor-intensive.}
\par{To constrain the synthesis of images and annotations at the same time without using a large scale of pre-labeling, we propose a novel controllable and simultaneous synthesizer (so-called CS$^2$) in this study. The novelty of our work is three-fold: 1) we develop a novel unsupervised mask-to-image synthesis pipeline that generates images controllably without human labeling; 2) instead of directly using the numeric and disarranged unsupervised segmentation masks, which are cluttered with over-segmented super-pixels, we assign the mean Hounsfield unit (HU) value for each cluster in the unsupervised segmentation masks to obtain an ordered and well-organized labeling; and 3) we propose a new synthesis network structure featured by multiple adaptive instance normalization (AdaIN) blocks that handles unaligned structural and tissue information. The code is publicly available at https://github.com/ayanglab/CS2. }
\section{Methods}
Our model consists of three major components, including 1) an \textbf{unsupervised mask generation} module, which generates unsupervised segmentation masks from input images; 2) a \textbf{multiple AdaIN GAN} that receives an unsupervised mask and a reference CT image as inputs, and outputs a synthetic CT image; then we use a small labeled subset of synthetic images and leverage the feature maps from the mask-to-image model to train 3) an \textbf{ensemble pixel-wise classifier for semantic segmentation}. Once trained, our model can produce synthetic images and corresponding annotations simultaneously. We present the overall network structure in Fig. \ref{fig:1} (d) and introduce the details of each component as follows.  

\begin{figure}
\includegraphics[width=\textwidth]{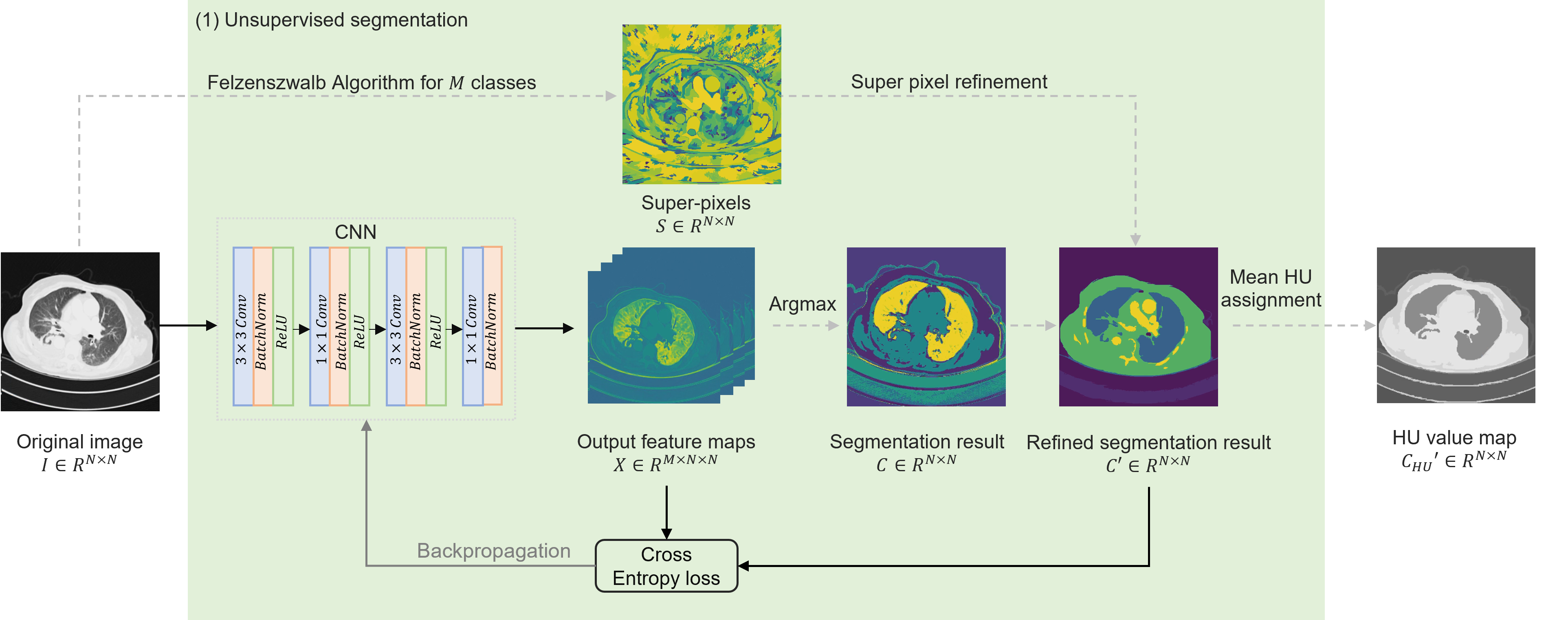}\vspace{-1em}
\caption {The workflow for unsupervised mask generation. Dashed lines indicate no gradient back-propagation during the training procedure. \vspace{-1em}} \label{fig:2}
\end{figure}

\subsection{Unsupervised Mask Generation}
 To achieve a controllable synthesis, an adjustable input that contains the structure of synthetic images must be provided. As we mentioned in the Introduction section, the supervised generation of structural masks requires a large scale of pre-labeling. Thus, in our algorithm, we adopt an unsupervised structural mask generation \cite{kanezaki2018unsupervised} as shown in Fig. \ref{fig:2}. Unsupervised segmentation masks are obtained by a super-pixel guided CNN. 
 
 Instead of using the disarrayed unsupervised masks directly, we compute the average HU value inside each class and assign this value to each cluster to create the structural guidance maps. In so doing, we can achieve an ordered and well-organized labeling for our unsupervised segmentation mask generation. We refer to this step as the mean HU assignment. According to the mean HU assignment, we can easily edit the unsupervised segmentation mask. For example, by adding a patch with an approximate HU value of the COVID infections, we can artificially add a lesion in the synthesized healthy lungs. It is of note that this mean HU assignment step is also crucial for the content matching loss defined in Section 2.2.
 
 The unsupervised segmentation algorithm is optimized by the cross-entropy loss between CNN output $X$ and the super-pixel refined mask $C'$. $M$ is the number of classes initialized for our algorithm and is a pre-defined super parameter. For each cluster $m$ in the super-pixel mask $S$, we extract the corresponding region $C^{(m)}$ in segmentation result $C$. Then we count the mostly appeared class in $C^{(m)}$, and assign this value $C_{\text{max}}$ as the new label to all elements in $C^{(m)}$ The elements in $C^{(m)}$ is mathematically defined as 
$$
  C_{i,j}^{(m)}=
\begin{cases}
0& S_{i,j}\neq m\\
1& S_{i,j}=m
\end{cases}
. 
$$
Here, $C_{\text{max}},m,C_{i,j},S_{i,j}$ are class numbers in the range of $(0,M]$. The refined mask $C'=\sum_{m=1}^M C^{(m)}$. This refinement can decrease the number of classes in the segmentation results. 
Once the number of clusters in the refined segmentation result reaches four or the number of iterations reaches 60, we stop the training of the CNN and use the refined segmentation result $C'$ as our unsupervised masks.
\subsection{Multiple AdaIN GAN} 
\begin{figure}[t]
\includegraphics[width=\textwidth]{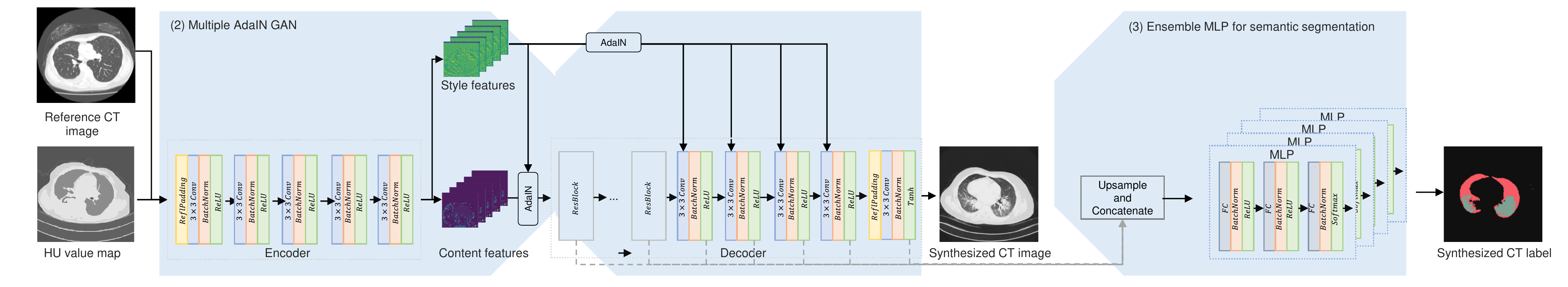}\vspace{-1em}
\caption {The network structure of our UM2I model and the ensemble MLP pixel-wise classifier for the segmentation synthesis. Dashed lines indicate no gradient back-propagation during training in this operation.\vspace{-1em}} \label{fig:3}
\end{figure}

For the image synthesis branch, we use a multiple AdaIN GAN shown in Fig. \ref{fig:3}. The input of our generator is one HU value map and one non-corresponding reference CT image. An additional reference CT image helps the synthesis network to adopt the style of the whole dataset better, and the unpaired inputs increase the variability of the synthetic images. The encoder extracts feature from both HU value maps and CT images simultaneously, and an Adaptive Instance Normalization (AdaIN) \cite{huang2017arbitrary} block normalize the content feature (from the HU value map) with the style feature (e.g., from the CT image). In our multiple AdaIN GAN, AdaIN blocks are added in all convolutional blocks except for the residual blocks (ResBlocks) \cite{he2016deep}. This is because AdaIN changes the distribution of feature maps; thus, it corrupts the identity mapping between input feature maps and output feature maps in residual blocks. 

The final loss function is composed of three parts: 1) a discriminator loss, which is computed by a patch-based discriminator \cite{demir2018patch}; 2) a style matching loss, which is the VGG loss \cite{johnson2016perceptual} between the reference CT and our synthesized CT images; and 3) a content matching loss that is the Mean Squared Error (MSE) loss between the content and our synthesized CT. 
In the previously published style transfer network using AdaIN \cite{huang2017arbitrary}, the content matching loss was computed by the VGG loss between content features and the feature maps extracted from the synthesized images. However, this loss function required extra GPU memory and limited the training batch size. We observed in our experiments that a large batch size led to a stable optimization procedure, which was also in accordance with a previous study \cite{brock2018large}. Therefore, we replace the content matching loss using the MSE between the HU value map and our synthesized CT images. Due to our mean HU assignment step, this MSE loss function performs better compared to using the original content matching loss. 

\subsection{Ensemble MLP Classifier for Semantic Segmentation}
Once the generative model is trained, we can generate synthetic images. The multiple AdaIN GAN provides effective image representations for synthetic images. We label several synthetic images (30 in our experiments) and then up-sample their features from the decoder into the same size as the original CT image. The features in the decoder are so powerful that even labeling a handful of synthetic images can lead to an accurate segmentation synthesis. We then train an ensemble multi-layer perceptron (MLP) classifier for the pixel-wise classification, as shown in Fig. \ref{fig:3}. The ensemble MLP classifier and multiple AdaIN GAN are trained separately. During inference, synthesized CT images and corresponding annotations are produced. In our experiments, we trained 10 MLP classifiers and obtained the final segmentation mask by majority voting. The number of pixel-wise classifiers was selected based on empirical studies according to \cite{zhang2021datasetgan}.

\section{Experiments}
\subsection{Dataset}
\noindent We used two datasets to validate the performance of our algorithm. The target of our experiment is to segment both lung tissues and ground glass opacities (GGOs) from CT images in those datasets. The first dataset is our in-house multi-center chest CT dataset, which contains 7,140 3D CT volumes from 2,071 COVID patients. An experienced radiologist manually labeled 1,192 volumes for this dataset. We randomly selected 192 volumes for independent testing. Detailed descriptions of our private dataset can be found in our supplementary file. The second is an open-access chest CT dataset, COVID-19-CT-Seg, containing 20 3D CT volumes of COVID patients \cite{ma2020towards}. All CT images from this open-access dataset have left lung, right lung, and infections labeled by two radiologists and verified by an experienced radiologist. It should be noted that we only used the open-access dataset for independent testing.

Our experiments were performed on 4-channel 2.5D CT images, i.e., we selected 4 2D slices from the 3D CT volume and stacked them channel-wise. The 2.5D concept \cite{roth2014new} is widely used to represent 3D medical imaging volumes and to reduce computational costs. In this study, we used 2.5D CT representation to handle the varied slice thickness among multi-centered CT data. The selection strategy was a uniform selection from the middle (25$\%$-75$\%$) slices. 

\subsection{Experimental Settings}
We compared our CS$^2$ with three state-of-the-art models, as shown in Fig. \ref{fig:1}. For vector-to-image (V2I) synthesis, we used StyleGAN \cite{karras2019style}. For mask-to-image synthesis (M2I), we used the pix2pix \cite{isola2017image} model, and the masks are generated by randomly adding circular GGO patches to lung segmentation masks. For the vector-to-mask-to-image synthesis (V2M2I), we used styleGAN to generate synthesis masks and then used the pix2pix model to generate patches. For each generative model, we created synthetic datasets with 4,000 2.5D CT images. We will refer to 2.5D CT images as images or samples in the following section. 

We further trained a UNet on these synthetic images together for lung and GGO segmentation. To prove that our synthetic data are representative, we also compared the segmentation result with the transfer-learning based method, i.e., we used a pretrained nnUNet model \cite{isensee2018no} (task numbered 115 \cite{roth2021rapid}) and fine-tuned the last layer according to 10 volumes from our labeled in-house training dataset. 

\subsection{Results}
\begin{figure}
\includegraphics[width=12 cm]{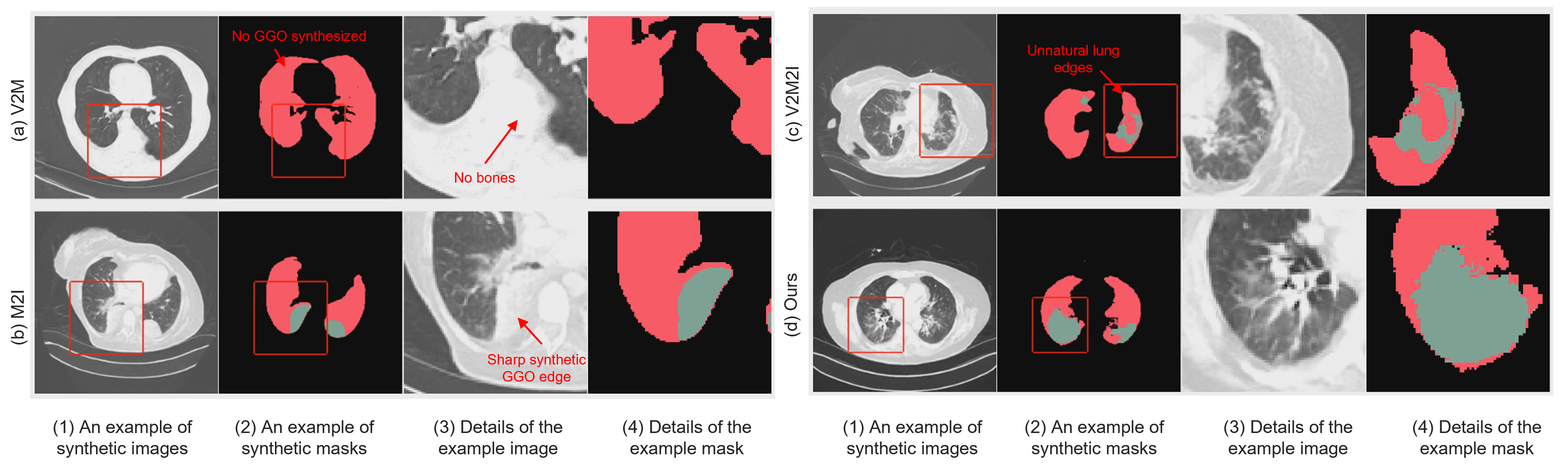}\vspace{-1em}
\caption {Example synthetic images from four generative models. The masks are synthetic segmentation masks that corresponded with these images. Red pixels indicate lung tissues and green pixels are GGOs. \textbf{We also include synthetic samples of different modalities in our supplementary file.}\vspace{-1em}} \label{fig:4}
\end{figure}
\noindent\textbf{Our model can successfully synthesize both the overall shape of lungs and textures of the GGOs.} Examples of synthetic images and annotations are shown in Fig. \ref{fig:4}. The V2I model, although generates a realistic overall shape of the lung and human body, falsely captures the texture of lung tissues, as shown in Fig. \ref{fig:4} (a1). Since the segmentation mask of the M2I model is manually generated, the M2I model failed to capture the gradual change between normal tissues and GGOs, as shown in Fig. \ref{fig:4} (b3). As for the V2M2I synthesized masks shown in Fig. \ref{fig:4} (c2), the lung edges are unnatural. Our model successfully generates realistic synthetic images and annotations. Both M2I and V2M2I models are trained with 500 2.5D CT images. 
\begin{figure}[b]
\includegraphics[width=\textwidth]{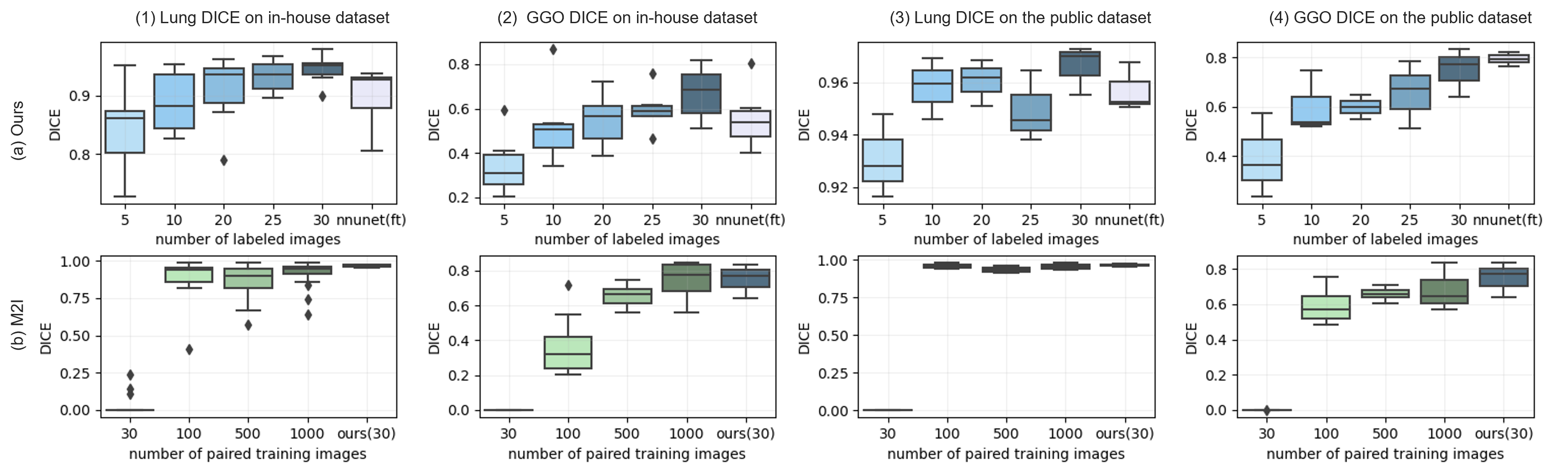}\vspace{-1em}
\caption {Dice scores of the lung and GGO on both in-house and public datasets from models trained on CS$^2$ synthesized (a) and M2I model synthesized (b) datasets.\vspace{-1em}} \label{fig:5}
\end{figure}

\textbf{The downstream segmentation model trained on CS$^2$ synthesized images can reach an accuracy as high as fully supervised pretrained nnUNet.}  We trained our ensemble MLP classifier with a different number of labeled synthetic 2.5D CT images and generated 4,000 synthetic masks for each group. Then we trained our segmentation network with these synthetic data. The result is shown in Fig. \ref{fig:5} (a). In our in-house dataset, the UNets trained on CS$^2$ synthetic datasets with 20, 25, and 30 manual labels even outperform nnUNet which was trained on 199 manually labeled 3D CT volumes and fine-tuned by 10 volumes in a fully supervised manner. It is of note that in the public dataset, the UNet trained on our synthetic datasets did not perform as well as the fine-tuned nnUNet. However, our segmentation results are still comparable in such a low-labeled data scenario. 

\textbf{Our CS$^2$ demonstrate promising results for large-scale annotated data synthesis when only a limited number of annotations is required.} However, when the number of annotations increased, the performance of the M2I model and V2M2I model improved, as shown in Fig. \ref{fig:5} (b). We only plotted the M2I model performance here because the performance of the V2M2I model was limited by its M2I component. To achieve similar synthesis performance, the M2I model required at least 500 paired training images.
\begin{figure}[t]
\includegraphics[width=\textwidth]{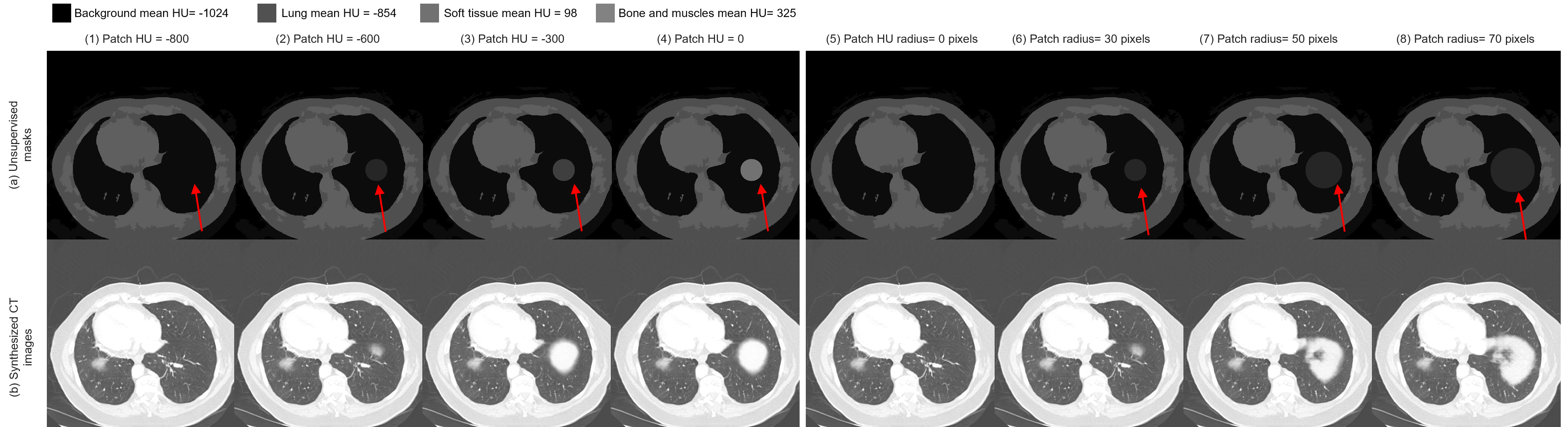}\vspace{-1em}
\caption {An example of our synthetic images (b) structurally edited with the Unsupervised masks (a) by adding circular patches of different HU values (1-4) and radii (5-8). The patches in (a1) to (a4) have a radius of 30 pixels, and the patches in (a1) to (a4) have a mean HU value of -600.\vspace{-1em}} \label{fig:6}
\end{figure}

\textbf{The synthetic images of our model are structurally editable}. As shown in Fig. \ref{fig:6}, we added circular patches in the lung masks with different mean HU values and different shapes. The mean HU values of the added patches are more decisive than their radius. For example in (b2) to (b4), even though we did not increase the radius of the added patches, our model detected a change from GGOs patches toward background tissue patches, thus automatically adjusting the size of the generated infectious areas. In contrast, as (b4) to (b8) show, once the mean HU value is fixed, the corresponding object is synthesized (e.g., no extraneous tissue or lesion is generated). The example HU values were given by an experienced radiologist, and we present the statistical distributions of HU values in unsupervised masks in our supplementary file.

\section{Conclusion}
In this study, a generative model called CS$^2$ has been proposed for data augmentation. By labeling only a handful of synthetic images (e.g., 30 2.5D CT images in our experimental studies), our CS$^2$ model has been able to generate realistic images and annotations at the same time. In addition, our algorithm has no requirement of any large pre-labeling or manually adjusted segmentation masks. Our experiments have proven our hypothesis that the proposed CS$^2$ has produced realistic synthetic images with annotations, which have enhanced the accuracy of lung tissues and infections segmentation of COVID-19 CT images. One limitation of our work is that we obtained our segmentation masks from a pixel-wise classifier, so the segmentation masks are fuzzy and have a few (~8$\%$) scattered wrongly labeled pixels. Post-processing algorithms on segmentation masks such as connected component analysis can remove these artifacts easily. 

\section{Acknowledgement}
This study was supported in part by the ERC IMI (101005122), the H2020 (952172), the MRC (MC/PC/21013), the Royal Society (IEC\textbackslash NSFC\textbackslash211235), the NVIDIA Academic Hardware Grant Program, the SABER project supported by Boehringer Ingelheim Ltd, and the UKRI Future Leaders Fellowship (MR/V023799/1).

%
%
%
%

\bibliographystyle{splncs04}
\bibliography{paper133.bib}

\appendix
\renewcommand\thefigure{\thesection\arabic{figure}}    
\renewcommand\thetable{\thesection\arabic{table}}
\section{Supplementary File}
\setcounter{figure}{0}   
\setcounter{table}{0} 
\begin{figure}
\includegraphics[width=\textwidth]{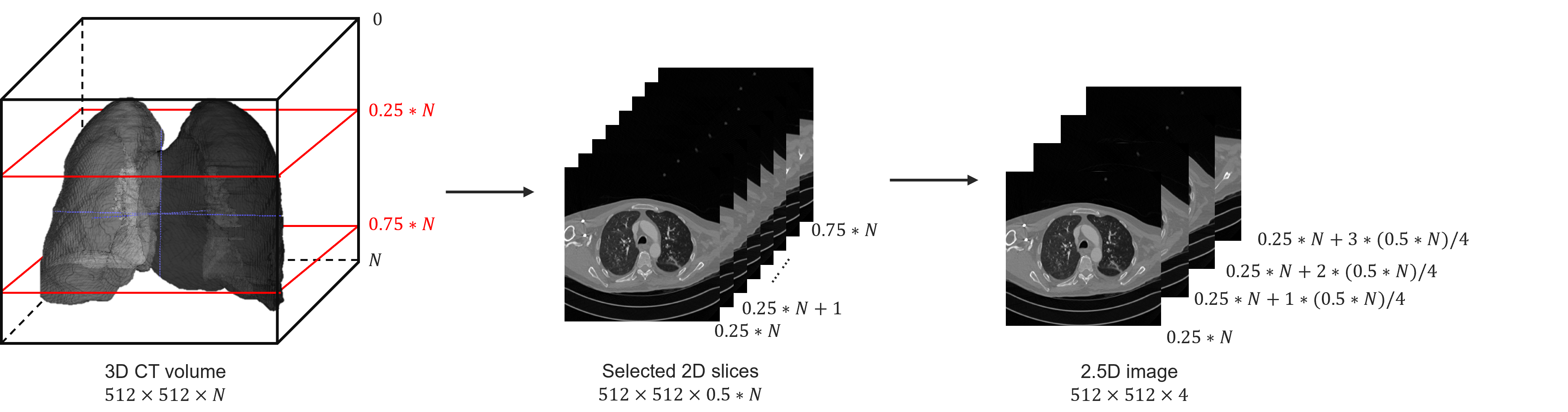}
\caption {The generation of our 4-channel 2.5D CT images. Here, N is the number of slices. Pixel spacing of these images is [0.6875 mm, 0.6875 mm], leading to an image array size of 512×512 for each 2D slice in these 3D volumes. The number of slices varies in a range of [36, 424]. We used a round-up strategy. For example, if the number of slices of one CT volume is N=31, the selected CT 2D slices are chosen from [7,23], and the four slices in this 2.5D CT case are the ones numbered 7,11,15 and 19.} \label{fig:S1}
\end{figure}

\begin{table}\centering
\caption{The architecture of models used in the manuscript.}
\label{tab2}
\begin{tabular}{ccc}
\hline
Model name& 	Architecture Used& 	Open-source code based on\\
\hline
V2I model&	StyleGAN&	https://github.com/huangzh13/StyleGAN.pytorch\\
\hline
M2I model&	Pix2pix&	https://github.com/NVIDIA/pix2pixHD\\
\hline
V2M2I model&	StyleGAN and pix2pix& \\
\hline
\end{tabular}
\end{table}

 \begin{figure}
\includegraphics[width=\textwidth]{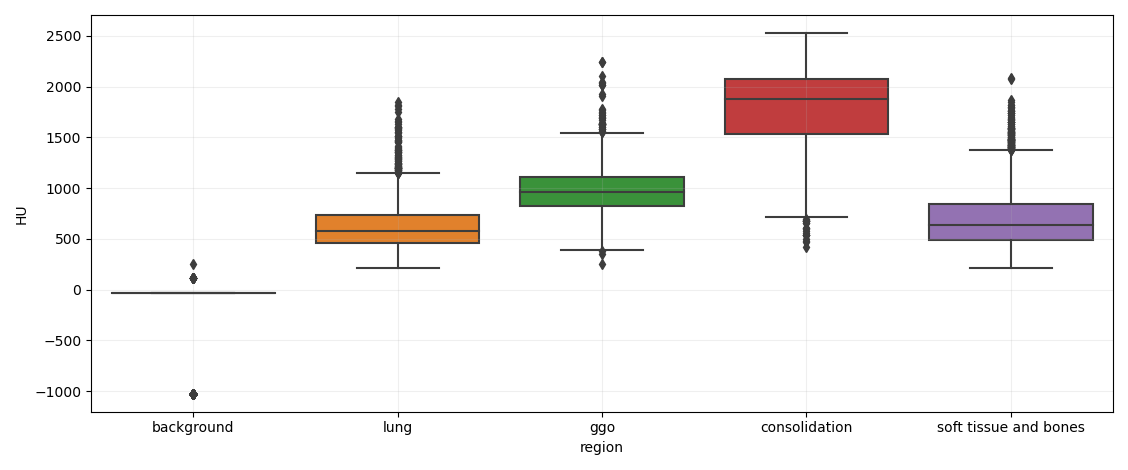}
\caption {We manually segmented 200 labeled CT images from our in-house dataset and calculate the distribution of HU values in three regions including background, lung, GGO, and other tissues. Our CS2 network can be structurally edited with the guidance of mean HU values.} \label{fig:S2}
\end{figure}

\begin{figure}
\includegraphics[width=\textwidth]{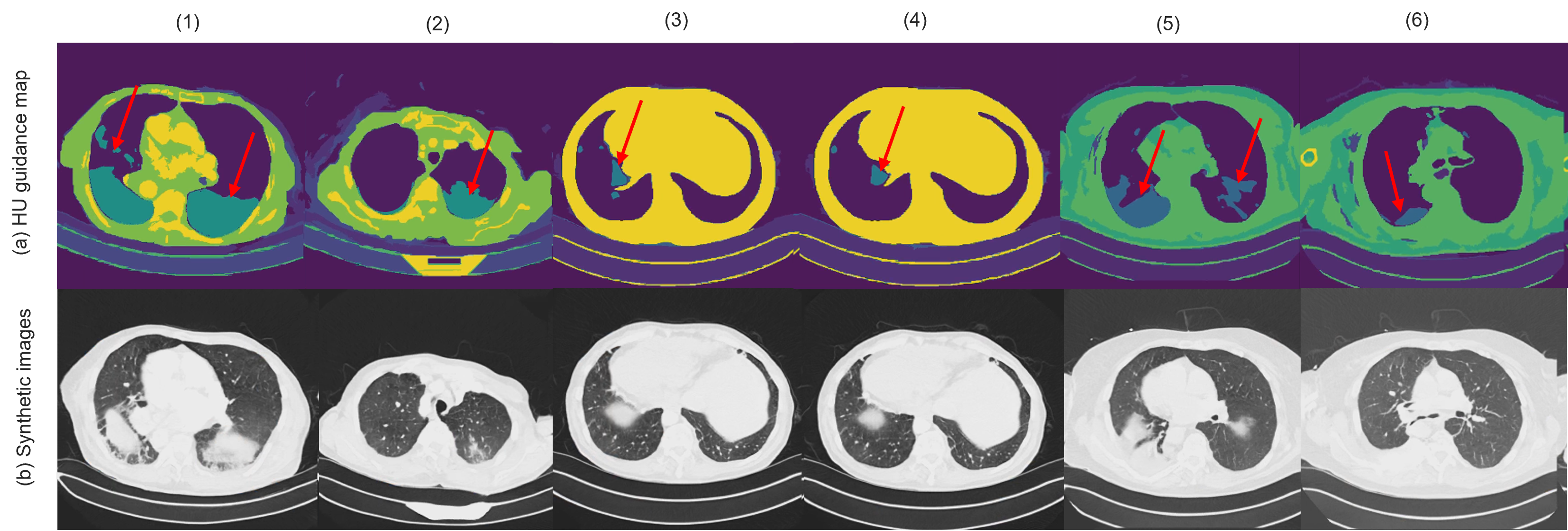}
\caption {Examples of synthetic images with manually added lesion patches. Despite of circular patches, we also added some manually drawn patches to visualize the synthesis result.} \label{fig:S3}
\end{figure}

\begin{figure}
\includegraphics[width=\textwidth]{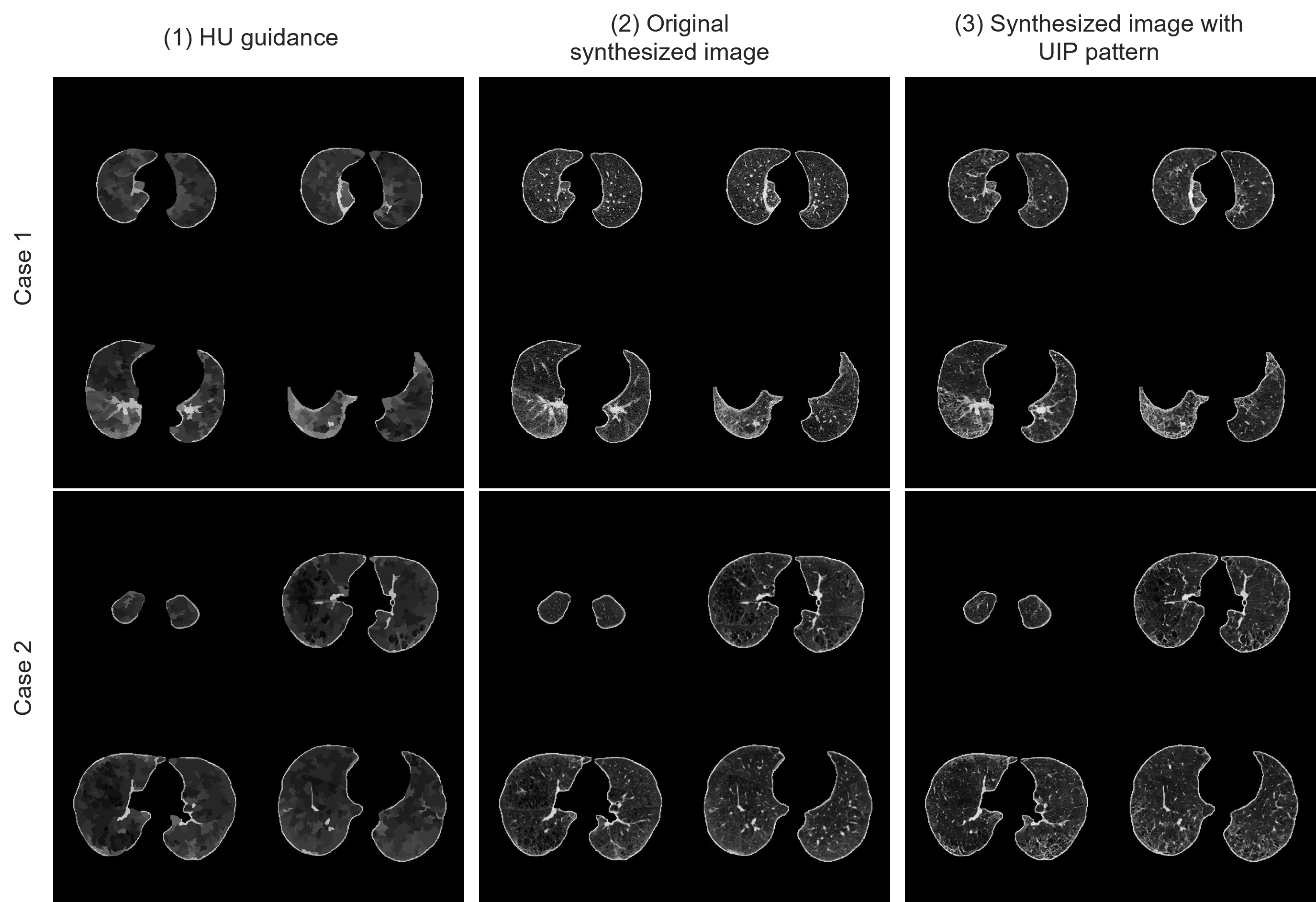}
\caption {Synthetic idiopathic pulmonary fibrosis (IPF) cases. The HRCT appearance of IPF is more complex than GGO and consolidation. We trained a synthesizer on UIP cases only, and inferenced the UIP synthesizer with non-UIP HRCT images. As is shown, by improving the class number of unsupervised guidance maps, we can successfully model the appearance of IPF as well as controllably synthesize IPF CT images.} \label{fig:S4}
\end{figure}
 
\begin{figure}
\includegraphics[width=\textwidth]{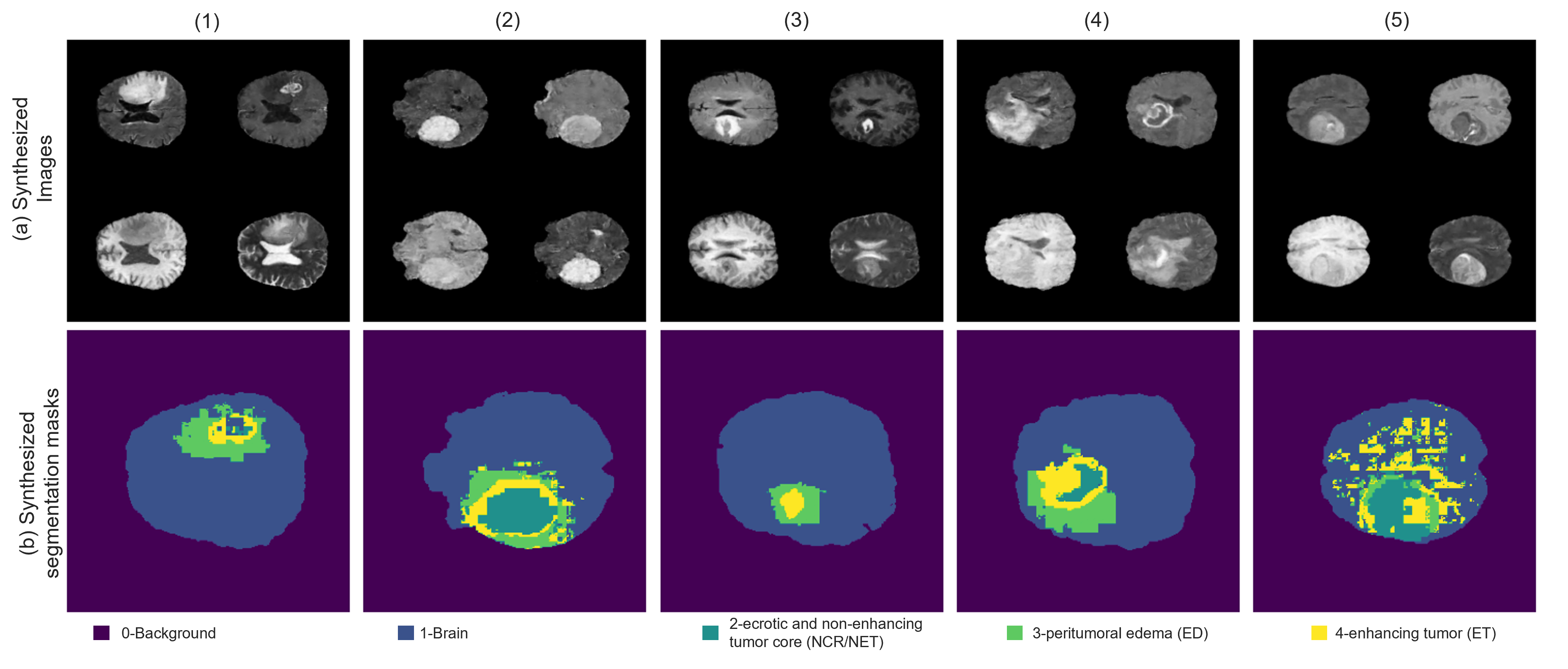}
\caption{ Synthetic brain tumor cases. Our V2M2I model can synthesize four modalities at the same time, as well as their segmentation results. Fig. 5 is a typical failed cases with scattering wrongly segmented pixels with 8$\%$ wrongly segmented pixels. Post-processing algorithms such as connected component analysis, can remove these scatters.  } \label{fig:S5}
\end{figure}

\end{document}